\documentclass[preprint]{JHEP3}

\usepackage{amssymb}
\usepackage{amsmath}
\usepackage{graphicx}

\newcommand{\ii}{\'\i} 
 
\def\beq{\begin{eqnarray}}    
\def\eeq{\end{eqnarray}}      

\def\Box{\square}                  
\def\tr{\,\mbox{tr}\,}                  
\def\Tr{\,\mbox{Tr}\,}                  

\def\al{\alpha}
\def\be{\beta}

\def\de{\delta}

\def\vp{\varepsilon}

\def\La{\Lambda}
\def\la{\lambda}
\def\na{\nabla}
\def\pa{\partial}

\def\om{\omega}
\def\ph{\varphi}

\def\Ga{\Gamma}

\def\La{\Lambda}

\title{Renormalization Group and Decoupling in Curved Space:
\\
III. $\,$ The Case of Spontaneous Symmetry Breaking}

\author{Eduard V. Gorbar 
\\
Bogolyubov Institute for Theoretical Physics, Kiev, Ukraine
\\ 
E-mail: \email{gorbar@bitp.kiev.ua}}

\author{Ilya L. Shapiro\thanks{On leave from Tomsk State 
Pedagogical University, Tomsk, Russia.}
\\
Departamento de F\ii sica, ICE, Universidade Federal de 
Juiz de Fora
\\

MG, Brazil
\\ 
E-mail: \email{shapiro@dftuz.unizar.es}}

\abstract{We continue investigation of the renormalization group 
and decoupling of the quantized massive fields in curved space
\cite{apco}. In the present work we analyze a theory, where 
fields gain their masses due to the Spontaneous Symmetry 
Breaking (SSB), that is the case providing a remarkable 
exception from the Appelquist-Carazzone theorem in the 
matter fields sector. In the vacuum sector, already at the 
classical level, the theory with SSB includes, in the general 
case an infinite number of the non-local terms in the induced 
vacuum action. Despite this surprising property, we show that 
the theory is renormalizable and moreover the low-energy 
decoupling in the higher-derivative gravitational sector 
performs similar to the AC theorem.}

\received{November, 2003}
\preprint{DF/UFJF-03/09}

\begin{document}

\section{Introduction}

\quad\quad
The renormalization group and decoupling of quantized massive
fields \cite{AC}
in curved space are important (but maybe not
always noticed) aspects of many modern theories of quantum
gravity. For example, the effective low-energy quantum gravity
\cite{don,tsamis} is based on the assumption that the decoupling
really takes place. This enables one to separate the quantum
effects of the heavy fields from the ones of the light fields and
in particular of gravitons. The decoupling is in the heart of the
cosmological applications \cite{nova,asta} of the semiclassical
approach to gravity (see, e.g. \cite{birdav,book}), where the
metric is considered as a classical external background for the
quantized matter fields. In a more general framework, the
low-energy spectrum of the (super)string theory includes
large amount of massive degrees of freedom and the consistency
of the low-energy predictions of string theory implies that the
virtual loops of these excitations do not affect the gravity
dynamics for the sole reason they have a large mass and decouple.
The standard Appelquist-Carazzone - like \cite{AC} form of
decoupling of the quantized massive fields in curved background
has been always anticipated \cite{don,tsamis,guber},
however the practical calculations of the decoupling have been
started only recently by the authors in \cite{apco}. In these
papers we have performed the calculations of the 1-loop Feynman
diagrams for the graviton propagator on the flat background
using a physical mass-dependent renormalization scheme, and
also equivalent covariant calculations in the second order
in the curvature tensor. In the higher derivative vacuum
sector we met an expected form of decoupling, similar to
the one of the Appelquist-Carazzone theorem in a matter
sector. Unfortunately, for the cosmological $\,\,\La\,\,$ and
inverse Newton $\,\,1/G\,\,$ constants one does not see the
decoupling and even the $\, \be$-functions themselves. The
most probable reason is the restricted power of the available
calculational methods, which are essentially based on the
perturbative expansion on the flat background or on the
equivalent covariant procedure. Indeed, we have to assume that
the $\be$-functions for $\,\,\La\,\,$ and $\,\,1/G\,\,$ exist,
for otherwise we would meet a disagreement between the
mass-dependent renormalization scheme and the completely
covariant minimal subtraction $\,\,{\overline{MS}}\,\,$
scheme where these two $\be$-functions
can be easily obtained \cite{book}. It is worth noticing
that the
simplest assumption concerning the form of decoupling
for the cosmological constant leads to the potentially
testable running of the cosmological constant in the late
Universe \cite{rgCC} and therefore this problem deserves
a special attention.

All the results of \cite{apco} concern the decoupling
of the massive fields in curved space-time. But there is
an interesting aspect of the decoupling which has to be
considered separately.
In many cases the fields which are massless in the initial
classical Lagrangian become massive due to the SSB mechanism
and we need to know whether the decoupling takes place,
also, for these fields. The special interest to this problem
is due to the well-known fact that in the matter field sector
the theories with the SSB violate the Appelquist-Carazzone
theorem (see, e.g. \cite{Collins}). One can, naturally, wonder
whether something like that happens or not in the case of gravity.
Perhaps, it is worth remembering that all massive particles
in the Standard Model gain their masses due to the SSB.
One can guess
that this is also the case for the Grand Unification Theories
at higher energies. Furthermore, the possibility of the
supersymmetry breaking due to the SSB can not be completely
ruled out \cite{susy}, and therefore the understanding of
decoupling in an external gravitational field in the theories
with SSB looks rather general problem.

The purpose of the present article is to clarify the issue
of decoupling in the SSB theories. The paper is organized as
follows. The section 2 is
devoted to the description of the SSB at the tree level
in the case of the non-minimal coupling of the scalar
(Higgs) field with gravity.
We find out that the low-energy induced gravitational
action includes, along with the Einstein-Hilbert term,
an infinite set of the non-local terms. In section 3
the general discussion of the one-loop corrections
to the induced gravitational action in the theory
with SSB and for the case of the minimal interaction
is given. In particular, the gauge-fixing invariance of
the quantum corrections in the theory with SSB is proved
explicitly.
In sections 4 and 5 we generalize these considerations
to the general situation with the non-minimal
interaction. In section 4 the logarithmic divergences
are calculated and the general scheme of renormalization
of the vacuum sector in the theory with SSB is
outlined. In particular, we show that the non-local
terms must be included into the vacuum action and that
these terms must be renormalized at quantum level.
In section 5 the physical mass-dependent scheme of
renormalization is applied to the SSB theory with
the general non-minimal coupling between scalar and
gravity. An explicit form of the $\,\be$-functions in
the vacuum sector are derived and the low energy
decoupling (analog of the Appelquist and Carazzone
theorem) is established for the parameters corresponding
to all higher derivative terms. Finally, we summarize the 
results in section 6.

\section{SSB and the non-local vacuum action}

In this paper we shall deal with the following
classical action of charged scalar $\,\ph\,$ coupled
to the Abelian gauge vector $\,A_\mu$:
\beq
S\,=\,\int d^4x\sqrt{-g}\,\Big\{
-\frac14\,F_{\mu\nu}\,F^{\mu\nu}\,+\,
g^{\mu\nu}\,(\pa_\mu-ieA_\mu)\ph^*\,(\pa_\nu+ieA_\mu)\ph\,+
\nonumber
\\
+\,\mu_0^2\,\ph^*\ph-\la(\ph^*\ph)^2
\,+\,\xi\,R\,\ph^*\ph\Big\}\,.
\label{SSB 1}
\eeq
The generalization for the non-Abelian theory
would be straightforward, but there is no reason to consider
it because we are interested in the one-loop vacuum effects
and the results are indeed the same for both Abelian and
non-Abelian cases.

Our first purpose is to investigate the SSB at the classical
level. The VEV for the scalar field is defined as a solution
of the equation
\beq
-\,{\Box}v\,+\,\mu_0^2 v\,+\,\xi R\,v\,-\,2\la v^3\,=\,0\,.
\label{SSB 2}
\eeq
If the interaction between scalar and metric is
minimal $\,\xi=0$, the SSB is standard and simple, because
the vacuum solution of the last equation is constant
\beq
v_0^2\,=\,\frac{\mu_0^2}{2\,\la}\,.
\label{SSB 3}
\eeq
In the last expression we have introduced a special notation
$\,v_0\,$ for the case of a minimal interaction, in order
to distinguish it from the solution $\,v\,$ of the general
equation (\ref{SSB 2}). Starting from (\ref{SSB 3}), the
conventional scheme of the SSB and the Higgs mechanism does
not require serious modifications because of the presence of
an external metric field. However, the consistency of the
quantum field theory in curved space requires the non-minimal
interaction such that $\,\xi\neq 0$ (see, e.g. \cite{book}
for the introduction).
For the general case of the non-constant
scalar curvature one meets, instead of Eq. (\ref{SSB 3}),
another solution $\,v(x)\neq \mbox{const}$. Hence, we
can not ignore the derivatives of $\,v\,$ and, unfortunately,
the solution for the VEV can not be obtained in a closed and
simple form.

Our main interest in this paper will be the decoupling of
massive fields at low energies, when the values of scalar
curvature are small. Therefore, we can try to consider
(\ref{SSB 3}) as the zero-order approximation and find the
solution of the Eq. (\ref{SSB 2}) in the form of the power
series in curvature
\beq
v(x)\,=\,v_0\,+\,v_1(x)\,+\,v_2(x)\,+\,...\,.
\label{SSB 4}
\eeq
For the first order term $\,v_1(x)\,$ we have the following
equation
\beq
-\,{\Box}v_1\,+\,\mu^2v_1\,+\,\xi R\,v_0
\,-\,6\la v_0^2\,v_1\,=\,0\,,
\label{SSB 5}
\eeq
and the solution has the form
\beq
v_1\,=\,\frac{\xi\,v_0}{{\Box}-\mu^2+6\la v_0^2}\, R
\,=\,\frac{\xi\, v_0}{{\Box}\,+\,4\la v_0^2}\,R\,,
\label{SSB 6}
\eeq
where we used (\ref{SSB 3}). In a similar way, we find
\beq
v_2\,=\,\frac{\xi^2\,v_0}{\Box + 4\la v_0^2}\,
R\,\, \frac{1}{\Box + 4\la v_0^2}\,R
\,-\,\frac{6\,\la\,\xi^2\,v^3_0}{\Box + 4\la v_0^2}\,
\Big(\,\frac{1}{\Box + 4\,\la v_0^2}\,R\,\Big)^2\,,
\label{SSB 7}
\eeq
where the operator in each parenthesis acts only
on the curvature inside this parenthesis. Contrary to that,
the left operator $\,[\Box + 4\la v_0^2]^{-1}\,$ in the
first term at the {\it r.h.s.} of the last equation
acts on all expression to the right of it. In general,
here and below the parenthesis restrict the action of
the differential or inverse differential (like
$\,[\Box + 4\la v_0^2]^{-1}$) operators. 

Of course,
one can continue the expansion of $\,v\,$ to any desirable
order. If we replace the SSB solution $v(x)$ back into 
the scalar section of the action (\ref{SSB 1}) we obtain 
the following result for the induced low-energy action 
of vacuum:
\beq
S_{ind}
\,=\,\int d^4x\sqrt{-g}\,\Big\{\,g^{\mu\nu}
\,\pa_\mu v \,\pa_\nu v \,+\,(\mu_0^2+\xi R)\,v^2
\,-\,\la\,v^4\,\Big\}\,.
\label{SSB 80}
\eeq

It is remarkable that, instead of the conventional
cosmological constant and Einstein-Hilbert term, here
we meet an infinite series of non-local expressions due 
to non-locality of (\ref{SSB 4}). Making an expansion
in curvature tensor, in the second order we obtain
\beq
S_{ind}
\,=\,\int d^4x\sqrt{-g}\,\Big\{-\,v_1{\Box}v_1\,+\,
\mu^2\,(v_0^2+2v_0v_1+2v_0v_2+v_1^2)
\nonumber
\\
-\,\la\,(v_0^4+4v_0^3v_1+4v_0^3v_2+6v_0^2v_1^2)
\,+\,\xi R\,(v_0^2+2v_0v_1)\,\Big\}\,+\,{\cal O}(R^3)\,.
\label{SSB 8}
\eeq
Now, using the equation (\ref{SSB 5}), after a small
algebra we arrive at the following action of induced
gravity
\beq
S_{ind}\,=\,\int d^4x\sqrt{-g}\,\Big\{
\,\la v_0^4\,+\,\xi Rv_0^2
\,+\,\xi^2\,v_0^2\,R\,\frac{1}{{\Box}
+4\,\la v_0^2}\,R\,+\,...\Big\}\,.
\label{SSB 9}
\eeq
The first term here is the induced cosmological constant,
which is supposed to almost cancel with its vacuum counterpart
(see, e.g. the discussion in \cite{nova}). The second term
is a usual induced Einstein-Hilbert action, which also
has to be summed up with the corresponding vacuum term.
Formally, both observables: the
cosmological and the Einstein-Hilbert terms, are given
by the sums of the vacuum and induced contributions.
However, there is a great difference between the two terms
from physical point of view. The observable cosmological
constant is extremely small compared to the magnitude of
the $\,\la v_0^4\,$, e.g. in the Standard Model of particle
physics. Hence there is an extremely precise cancelation
between the vacuum and induced cosmological constants
(see, e.g. \cite{weinRMP} for the introduction to the
cosmological constant problem and also \cite{nova} for
the discussion of the possible quantum effects).
At the same time, the situation for the 
Einstein-Hilbert term is quite different.
The overall coefficient here is nothing but the inverse
Newton constant $\,1/16\pi G\,=\,M_P^2/16\pi$, where
$\,M_P\approx 10^{19}\,GeV\,$ is a Planck mass. Of
course, the magnitude of this quantity is huge compared to
the induced term. The exception is indeed possible if we
assume a SSB phenomenon at the Planck scale, but at the
lower energies one can not distinguish the difference
between this ``induced'' gravitational action and the
``original'' vacuum one. Therefore, at low energies the
local induced Einstein-Hilbert term is irrelevant compared
to the classical (vacuum) gravitational action \footnote{The
remarkable exception is the possibility to have totally 
induced Einstein-Hilbert term (see, e.g. \cite{adler}
and references therein). 
Recently, the model of induced gravity found interesting
applications in the black hole physics \cite{Frolov}.}.

Besides the usual local terms, the induced tree-level
gravitational action includes an infinite set of the
non-local terms. These terms are somehow similar to the
nonlocalities which have been recently discussed
in \cite{dvali} in relation to the higher derivative
theories and the cosmological constant problem.
The appearance
of the non-local terms in the induced action (\ref{SSB 9})
is remarkable, also, for other reasons. Although the
coefficients of these terms are very small compared to the
vacuum Einstein-Hilbert term, the non-localities do not
mix with the local terms and, in principle, can lead to
some physical effects. If we consider the low energy SSB
phenomena in the framework of the SM, the non-local 
terms are irrelevant at low energies due to the large 
value of the
mass. But, if we assume that there is an extremely light
scalar (e.g. quintessence), whose mass is of the order
of the Hubble parameter and which has a potential admitting
a SSB, then the non-localities may become relevant and
in particular lead to observable consequences. In the next
sections we will not discuss these issues and will instead
concentrate on the quantum one-loop corrections to the
induced action (\ref{SSB 9}).

\section{The minimal interaction case and gauge fixing
independence}

In the previous section, we considered the SSB in curved 
space-time at the classical level. The next problem is to 
derive the quantum
corrections to the vacuum action from the theory (\ref{SSB 1})
with the SSB. Let us notice that the relevant form of the
contributions of the
massive scalar, fermion and vector fields to the effective
action of vacuum were already calculated in \cite{apco}.
These calculations enable one to see the decoupling of
massive fields at low energies through the application of
the physical mass-dependent renormalization scheme. The
methods used in \cite{apco} are based on the expansion of
the metric over the flat background or an equivalent
covariant expansion
of the effective action in the power series in curvature. Here
we shall generalize the same approach for the case when the
masses emerge as a result of the SSB in curved space-time.

Let us start from the simplest case of the
minimal interaction between scalar field and metric $\,\xi=0$.
The effective action $\,\Ga[\ph,g_{\mu\nu}]\,$ of the scalar
field can be presented as the perturbative expansion
\beq
\Ga[\ph,g_{\mu\nu}]\,=\,S_{cl}[\ph,g_{\mu\nu}]\,+\,
\hbar\,{\bar \Ga}^{(1)}[\ph,g_{\mu\nu}]\,+\,{\cal O}(\hbar^2)\,.
\label{3.1}
\eeq
In this paper we restrict the consideration by the one-loop
order and therefore we shall consider only the
$\,{\bar \Ga}^{(1)}[\ph,g_{\mu\nu}]\,$ term. Then the
effective equation for the VEV of scalar field has the form
\beq
\frac{\de S_{cl}}{\de \ph}\,+\,\hbar\,
\frac{\de {\bar \Ga}^{(1)}}{\de \ph}\,=\,0\,.
\label{3.2}
\eeq
Since $\,S_{cl}\,$ is given by the scalar sector of
(\ref{SSB 1}), the equation (\ref{3.2}) can be rewritten as
\beq
-\,{\Box}\ph\,+\,\mu^2\ph\,-\,2\la (\ph^*\ph)\ph\,+\,
\hbar\,\frac{\de {\bar \Ga}^{(1)}}{\de \ph}\,=\,0\,.
\label{3.3}
\eeq
Let us now present the scalar field as $\,\ph=v+\hbar \phi\,$,
where $\,v\,$ is the solution of the classical equation
$\,\,{\de S_{cl}}/{\de \ph}=0\,\,$ and $\,\hbar\phi\,$ is
a quantum correction. Then we find, in the first order in
$\,\hbar$, the following relation:
\beq
-\,{\Box}\phi\,+\,\mu^2\phi\,-\,6\la v^2\phi\,+\,
\hbar\,\frac{\de {\bar \Ga}^{(1)}[v,g]}{\de v}\,=\,0\,.
\label{3.4}
\eeq
After performing the expansion in $\,\hbar$, we find
\beq
\Ga[\ph,g_{\mu\nu}]&=&S_{cl}[v+\hbar\phi,\,g_{\mu\nu}]\,+\,
\hbar\,{\bar \Ga}^{(1)}[v+\phi,\,g_{\mu\nu}]\,+\,...
\nonumber
\\
\nonumber
\\
&= &
S_{cl}[v,\,g_{\mu\nu}]\,+\,\hbar\phi\,
\frac{\de S_{cl}[v,g]}{\de v}\,+\,
\hbar\,{\bar \Ga}^{(1)}[v,\,g_{\mu\nu}]\,+\,{\cal O}(\hbar^2)\,.
\label{3.5}
\eeq
Taking into account the equation of motion
$\,\,{\de S_{cl}(v,g)}/{\de v}=0$, we arrive at the
useful formula
\beq
\Ga[v+\hbar\phi,\,g_{\mu\nu}]\,=\,S_{cl}[v,\,g_{\mu\nu}]\,+\,
\hbar\,{\bar \Ga}^{(1)}[v,\,g_{\mu\nu}]\,+\,{\cal O}(\hbar^2)\,.
\label{3.6}
\eeq
The last relation holds even for the non-minimal scalar field,
and we shall use it in what follows. The equation (\ref{3.6})
shows that at the one-loop level one can derive the effective
action as a functional of the classical VEV. In the minimally
interacting theory this VEV is just a constant, but in the
general non-minimal case the classical VEV itself is a
complicated expression (\ref{SSB 4}).

Consider the SSB in the theory (\ref{SSB 1}) with $\,\xi=0$.
For this end we define $\,\ph=v+h+i\eta$ and replace it back
to the action. As far as we are interested in the one-loop
effects, we can keep the terms of the second order in the
quantum fields $\,\,h$, $\,\,\eta\,\,$ and disregard higher
order terms. In this way we arrive at the expression
for the quadratic in quantum fields part of the action
\beq
S^{(2)}=\int d^4x\sqrt{-g}\Big\{
(\pa_\mu h)^2
+ (\pa_\mu \eta)^2 - \frac14 F^2_{\mu\nu} + 2evA_\mu \na^\mu \eta
+ e^2v^2 A_\mu A^\mu - 4\la v^2 h^2\Big\},
\label{3.7}
\eeq
where we used notation
$\,(\pa h)^2=g^{\mu\nu}\pa_\mu h \pa_\nu h\,\,$. $\,$
Let us introduce the 'tHooft gauge fixing condition, depending
on an arbitrary parameter $\,\al$
\beq
S_{GF}\,=\,-\,\frac{1}{2\al}\,\int d^4x\sqrt{-g}\,
(\na_\mu A^\mu\,-\,2\,\al\,ev\,\eta)^2\,.
\label{3.8}
\eeq
Summing up the two terms we arrive at the expression for
the action with the gauge fixing term
\beq
S^{(2)}\,+\,S_{GF}\,=\,\int d^4x\sqrt{-g}\Big\{\,
-\frac14 F^2_{\mu\nu}\,-\,\frac{1}{2\al}\,(\pa_\mu A^\mu)^2
\,+\,e^2v^2 A_\mu A^\mu 
\nonumber
\\
+\,(\pa_\mu h)^2\,+\,(\pa_\mu \eta)^2\,
-\,4\la v^2h^2 \,-\,2\,\al e^2v^2 \eta^2\,\Big\}+...\,,
\label{3.9}
\eeq
where we kept only the terms of the second order in the
quantum fields $\,A^\mu , \,h , \,\eta$.

The action of the Faddeev-Popov gauge ghosts can be obtained
in a standard way as
\beq
S_{GH}\,=\,\int d^4x\sqrt{-g}\,{\bar C}
\,\left(\,\Box\,+\,2\,\al e^2v^2\,\right)\,C\,.
\label{3.10}
\eeq
After all, from the equations (\ref{3.9}) and (\ref{3.10})
we find that the one-loop corrections to the vacuum effective
action are given by the contributions of the fields
$\,A_\mu,\,$ $\,h,\,$ $\,\eta\,$, $\,{\bar C}\,$ and $\,C$
\beq
{\bar \Ga}^{(1)}[\ph,g_{\mu\nu}]\,=\,
\frac{i}{2}\Tr\ln\,\left[\,\de^\mu_\nu\,\Box\,-\,
\Big( 1-\frac1\al \Big)\na^\mu\na_\nu\,-\,R^\mu_{\,\nu}
\,+\,2e^2v^2\,\de^\mu_\nu\,\right]
\nonumber
\\
+\,\frac{i}{2}\Tr\ln\,(\,\Box \,+\,4\la v^2\,)\,
+\,\frac{i}{2}\Tr\ln\,(\,\Box \,+\,2\al\,e^2v^2\,)\,
-\,i\,\Tr\ln\,(\Box\,+\,2\,\al e^2v^2)\,.
\label{3.11}
\eeq
In the general case of an arbitrary $\,\al\,$ the first
term of the last expression is related to the functional
determinant of a non-minimal massive vector field.
This kind of operator have been never elaborated in
the literature\footnote{The reader can easily evaluate the
contribution of this operator to the UV divergences using
the consideration of
the rest of this section.}, moreover for the particular
value $\,\al=0\,$ the expression (\ref{3.11}) includes the
contributions of several massless modes, jeopardizing the
expected low-energy
decoupling. For all other values of $\,\al\,$ all the
degrees of freedom in (\ref{3.11}) are massive. Furthermore,
in the particular case $\al=1$ the above expressions
have only the well-known contributions of the minimal massive
vector and scalars. Indeed, the $\,\be$-functions for both
these cases were calculated in \cite{apco}. Then we can just
use the result of \cite{apco} where we have demonstrated the
universality of the decoupling of the massive fields at
low energies. Therefore, the decoupling is guaranteed if we
can prove the gauge-fixing independence of the effective
action.

Let us remind that there are general theorems concerning
the on-shell gauge independence. These theorems should be
directly applicable in our case because we are interested
in the vacuum effects which are not related to the equations
of motion for the matter fields. However, since the gauge
fixing independence has special importance here, it is worth
verifying it explicitly, at least for the particular case
$\,\xi=0$. The methods for investigating the
gauge fixing dependence have been developed in \cite{bavi},
and in this section we shall apply the modified version
of these methods for the
case of minimal coupling. As far as the gauge-fixing
independence is established, the derivation of the vacuum
$\,\be$-functions in the theory with SSB can be easily
performed using Eq. (\ref{3.11}) and the results of
\cite{apco}. The explicit calculation will be postponed for
the next section, where we shall consider a more general
theory with an arbitrary $\,\xi$.

Following \cite{bavi}, we shall evaluate the difference
between the Euclidean one-loop correction with an arbitrary
value of the gauge parameter $\,\,\al\,\,$ and the same
correction with the same parameter fixed $\,\,\al=1$.
\beq
{\bar \Ga}^{(1)}[\ph,g_{\mu\nu};\,\al]\,-\,
{\bar \Ga}^{(1)}[\ph,g_{\mu\nu};\,1].
\label{3.12}
\eeq
Let us start from the first term in Eq. (\ref{3.11}) and
take
\beq
{\hat {\cal F}}(\al)\,=\,{\cal F}_\mu^\nu(\al)\,=\,
\de_\mu^\nu\,\Box\,-\,
\Big( 1-\frac{1}{\al} \Big)\,\na_\mu\na^\nu\,-\,R_\mu^\nu
\,+\,m^2\,\de_\mu^\nu\,,
\label{3.13}
\eeq
where we denoted $\,m^2=2e^2v^2$. Consider the
difference
\beq
-\frac12\,\Tr\,\ln\,{\hat {\cal F}}(\al)\,+\,
\frac12\,\Tr\,\ln\,{\hat {\cal F}}(1)
\nonumber
\\
=\,-\frac12\,\Tr\,\ln\,\left[\,\de_\mu^\nu\,-\,
\Big( 1-\frac{1}{\al} \Big)\,\na_\mu\na^\nu\,
\frac{1}{\Box+m^2-R_{..}}\right]\,.
\label{3.14}
\eeq
Let us use the identity, derived in \cite{bavi} for an
arbitrary vector field $\,A_\nu$
\beq
\Big(\,\na^\mu\,\frac{1}{\Box-R_{..}}
\,-\,\frac{1}{\Box}\,\na^\mu\,\Big)\,A_\mu\,=\,0\,.
\label{3.15}
\eeq
The calculation of gauge fixing dependence performed in
\cite{bavi} concerns the massless gauge field, and in our
case we have a massive field. That is why we need to
generalize the identity (\ref{3.15}) for the massive case
\beq
\Big(\,\na^\mu\,\frac{1}{\Box+m^2-R_{..}}
\,-\,\frac{1}{\Box+m^2}\,\na^\mu\,\Big)\,A_\mu\,=\,0\,.
\label{3.16}
\eeq
This generalization can be performed by expanding the
identity (\ref{3.16}) into the series in $\,m^2$. First
we present the propagator for the massive case as a
series
\beq
\frac{1}{\Box + m^2 - R_{..}}
\,=\,
\sum_{n=0}^{\infty}(-1)^n
\,\frac{(m^2)^n}{(\Box-R_{..})^{n+1}}\,.
\label{3.17}
\eeq
The zero-order part of (\ref{3.17})
is the original identity (\ref{3.15}).
In the first order in $\,m^2\,$ we need to prove the
identity
\beq
\Big[\,\na^\mu\,\frac{1}{(\Box-R_{..})^2}
\,-\,\frac{1}{\Box^2}\,\na^\mu\,\Big]\,A_\mu\,=\,0\,.
\label{3.18}
\eeq
This can be easily done by presenting it in the form
\beq
\Big(\,\na^\mu\,\frac{1}{\Box-R_{..}}
\,-\,\frac{1}{\Box}\,\na^\mu\,\Big)\,\frac{1}{\Box-R_{..}}
\,A_\mu\,\,+\,\,\frac{1}{\Box}\,
\Big(\,\na^\mu\,\frac{1}{\Box-R_{..}}
\,-\,\frac{1}{\Box}\,\na^\mu\,\Big)\,A_\mu\,=\,0\,,
\label{3.19}
\eeq
where we have used the fact that the vector
$\,A^\prime_\mu\,=\,(\Box-R_{..})^{-1}\,A_\mu\,$ also
satisfies the identity (\ref{3.15}). The same operation
can be applied at any order in $\,m^2$, therefore
we proved the identity (\ref{3.16}).

Using this identity, one can rewrite the difference
(\ref{3.14}) as
\beq
-\frac12\,\Tr\,\ln\,\left[\,\de_\mu^\nu
\,-\,\Big( 1-\frac{1}{\al} \Big)\,\na_\mu\,
\frac{1}{\Box+m^2}\,\na^\nu\,\right]\,=
\\
=\,\frac12\,\Tr\,\sum_{n=1}^\infty\,
\frac{1}{n}\,\Big( 1-\frac{1}{\al} \Big)^n
\Big(\,\na_\mu\,\frac{1}{\Box+m^2}\,\na^\nu\,\Big)^n\,.
\label{3.20}
\eeq
Using the relation $\,\Tr(A\cdot B\cdot C)=\tr(C\cdot A\cdot B)$
and taking the trace over the indices $\,\mu\,$ and $\,\nu$,
after a small algebra, (\ref{3.20}) can be transformed into the
expression involving only the scalar operators (here and below
we disregard the $\,\Tr\ln\,$ of constants which are not relevant,
e.g. in the dimensional regularization)
\beq
-\frac12\,\Tr\,\left[\ln\,{\hat {\cal F}}(\al)\,-\,
\ln\,{\hat {\cal F}}(1)\right]\,=\,
-\frac12\,\Tr\,\ln\,\left(\,
\frac{\Box\,+\,\al m^2}{\Box\,+\, m^2}\,\right)\,.
\label{3.21}
\eeq
The gauge dependent part of the contribution of the
Higgs scalar is zero because the Higgs mass $\,M_H\,$
does not depend on the parameter $\,\al$. The gauge
dependent part of the contribution of the Goldstone
scalar (\ref{3.11}) is exactly the same as the
vector counterpart (\ref{3.21})
\beq
\frac{1}{2}\Tr\ln\,\,\left(\,
\frac{\,\Box \,+\,2\al\,e^2v^2}{\,\Box \,+\,2\,e^2v^2}
\,\right)\,=\,
-\frac{1}{2}\,\Tr\,\ln\,\left(\,
\frac{\Box\,+\,\al m^2}{\Box\,+\, m^2}\,\right)\,.
\label{3.22}
\eeq
Finally, the difference between the two ghost operators
(\ref{3.8}) contributes as
\beq
\Tr\,\ln\,\left(\,
\frac{\Box\,+\,\al m^2}{\Box\,+\, m^2}\,\right)\,.
\label{3.23}
\eeq
In total, three contributions (\ref{3.21}), (\ref{3.22}) and
(\ref{3.23}) give zero, and therefore the one-loop part of the
vacuum effective action in the theory with SSB is gauge
fixing independent. As it was shown above, this also means that
in the vacuum sector the quantum effects of the theory with the
SSB satisfy the Appelquist and Carazzone theorem \cite{AC} and
manifest the decoupling at low energies. In this respect the
gravitational vacuum quantum effects are very different from
the quantum effects in the matter sector where the mentioned
theorem may be violated for the masses of the SSB origin
(see, e.g. \cite{Collins}).


\section{$\overline{\bf MS}$-scheme
renormalization in the non-minimal case}

Let us first derive the one-loop divergences in the theory
with the SSB, using the background field method and the
Schwinger-DeWitt technique \cite{DeWitt65}. Starting
from the action (\ref{SSB 1}), we are going to integrate
over the matter fields $\,\ph ,\,A_\mu\,$ on the background
of the classical metric $\,g_{\mu\nu}$. In the spontaneously
broken phase the scalar field $\,\ph\,$ takes the VEV
$\,v\,$ corresponding to the solution of the equation
(\ref{SSB 2}). Indeed, this solution (\ref{SSB 4}) depends
exclusively on the metric. Therefore, we perform
the background shift of the scalar variable according to
\beq
\ph\,=\,v\,+\,h+i\eta\,,
\label{3-1}
\eeq
where $\,h\,$ and $\,\eta\,$ are real scalar quantum fields
(Higgs and Goldstone). Hence, we face a problem of deriving
the divergences in the theory with quantum fields $\,A_\mu$,
$\,h$, $\,\eta$, while the background fields include metric
and $\,v$, which, in turn, also depends on the metric.
As we shall see in what follows, the renormalization of the
theory looks rather standard in terms of $\,g_{\mu\nu}\,$
and $\,v$. However, it looks very unusual if we take the
expression (\ref{SSB 4}) for $\,v\,$ into account and 
express the effective action in the terms of metric.

In order to derive the one-loop quantum correction one needs
the part of the action which is bilinear in quantum fields.
Elementary calculations give the following result for the
sum of the action (\ref{SSB 1}) and the gauge-fixing term
(\ref{3.8})$\,$:
\beq
S^{(2)}\,+\,S_{GF}&=&\int d^4x\sqrt{-g}\Big\{\,
\frac12\,A_\mu \square A^\mu \,+\,
\frac{1}{2}\,\Big(\,1-\frac{1}{\al}\,\Big)\,
(\na_\mu A^\mu)^2\,-\,A^\mu A^\nu R_{\mu\nu}
\nonumber
\\
&+ & \frac12\,M_A^2A^2 +
(\pa_\mu h)^2+(\pa_\mu \eta)^2
-M_H^2h^2 -M_\eta^2\eta^2
-2\al e\eta\, A^\mu (\pa_\mu v)\Big\}\,,
\label{back 1}
\eeq
where we introduced new notations
\beq
M_A^2 = 2e^2v^2\,,\qquad
M_h^2 = 6\la v^2 - \mu_0^2 - \xi R\,,\qquad
M_\eta^2 = 2e^2v^2 + 2 \la v^2 -\mu_0^2 -\xi R\,.
\label{back 2}
\eeq
One can rewrite these quantities in a more
useful way. First we introduce
\beq
\xi{\cal K}=2\la\,\left(v^2-v_0^2\right)
=2\la v^2 -\mu_0^2\,.
\label{back 24}
\eeq
After replacing (\ref{SSB 4}), (\ref{SSB 6}) and
(\ref{SSB 7}) into (\ref{back 24}), in the lowest
order in curvature we obtain
\beq
\xi{\cal K} \,=\,\frac{2\xi v_0^2}{\Box+4\la v_0^2}\,R
\,+\,{\cal O}(R^2)\,.
\label{back 222}
\eeq
If we are interested in the low-energy effect, then the
derivatives of curvature are very small compared to
$\,v_0^2$. Then we have to expand the Green function,
in the expression above, as follows
\beq
\frac{1}{\Box+4\la v_0^2}\,=\,\frac{1}{4\la v_0^2}\,
\left( 1 - \frac{\Box}{4\la v_0^2}+\,...\right)
\,+\,{\cal O}(\Box R)\,.
\label{back 212}
\eeq
In the low-energy approximation we arrive
at the representation
\beq
\xi{\cal K} \,=\,\xi R
\,+\,\frac{\mbox{higher derivative terms}}{v_0^2}\,.
\label{back 232}
\eeq
Furthermore, the $\,(\square v)/v\,$ term admits the
following representation in terms of $\,{\cal K} \,$:
\beq
\frac{(\square v)}{v}
&=& \frac{\mu_0^2\,v+\xi R\,v-2\la\, v^3}{v}
= \xi R + 2\la v_0^2 - 2\la v^2
\,=\, \xi R - \xi {\cal K} \,.
\label{back 23}
\eeq

In the new notations
the elements of the expansion (\ref{back 2})
may be written in the form
\beq
M_A^2 &=& m^2 \,+\, \frac{e^2}{\la}\,\xi {\cal K} \,,
\,\,\qquad \qquad \qquad\qquad\,\,\,\,
m^2\,=\,2e^2v_0^2\,;
\nonumber
\\
M_h^2 &=& m^2_h\,-\,\xi\,R\,+\,3\,\xi\,{\cal K} \,,
\qquad\qquad\qquad\,\,\,
m^2_h\,=\,4\la v_0^2\,;
\nonumber
\\
M_\eta^2 &=& m^2 \,-\,\xi\,R\,+\,
\Big(\frac{e^2}{\la}+1\Big)\,\xi {\cal K} \,,
\label{back 22}
\eeq
where $\,m\,$ and $\,m_h\,$ are the
masses of the fields after SSB. Indeed, their values
are the same as in the minimal $\,\xi=0\,$ case.

Coming back to the calculation of divergencies,
since the one-loop effective action is gauge-fixing
invariant, (see the discussion in the previous section)
we can put $\,\al=1\,$. Then, making a
change of variables
\beq
h\, = \,\frac{i}{\sqrt{2}}\,\tilde{h}
\,,\qquad
\eta\, = \,\frac{i}{\sqrt{2}}\,\tilde{\eta}\,,
\nonumber
\eeq
we arrive at the following useful form of the
bilinear part of the action (\ref{back 2}):
\beq
S^{(2)}\,+\, S_{GF} &=& \int d^4x\sqrt{-g}\,\,\Big\{\,
\frac12\,\tilde{h}\,\hat{\cal H}_h\,\tilde{h}
\,\,\,+\,\,\,
\frac12\,\left(A^\mu \,\,\,\tilde{\eta}\right)\,
\hat{\cal H}\,
\left(\begin{array}{cc}
A_\nu \\ \tilde{\eta} \end{array}
\right)\,\Big\}\,,
\label{back 3}
\eeq
where the operators have the form
\beq
\hat{\cal H}_h \,=\, \square + M_h^2
\qquad \mbox{and} \qquad
\hat{\cal H}  \,=\,
\left\| \begin{array}{cc}\de^\nu_\mu \square - R^\nu_\mu
+ M_A^2 \de^\nu_\mu & -i\sqrt{2}\,e\,(\pa_\mu v) \\
-i\sqrt{2}\,e\,(\na^\nu v) & \square + M_\eta^2
\end{array}\right\|\,.
\label{back 4}
\eeq

Both operators $\,\hat{\cal H}_h\,$ and
$\,\hat{\cal H}\,$ have the standard structure
$\,\hat{1}\square + \hat{\Pi}\,$ and the algorithm for 
the divergences is well known (notice that the
masses are included into the operators $\Pi$ for
all the fields)
$$
\left.
\frac{i}{2}\,\Tr \ln (\hat{1}\square + \hat{\Pi})
\right|_{div}\,=
\,\frac{1}{2(4\pi)^2(2-\om)}\,\int d^4x\,\sqrt{-g}\,\tr\,
\Big\{\,\frac{\hat{1}}{180}\,\Big(
\,R_{\mu\nu\al\be}^2-R_{\mu\nu}^2+\square R\,\Big)
$$
\beq
+ \,\, \frac12\,{\hat P}^2
\,+\,\frac{1}{12}\hat{S}_{\mu\nu}\hat{S}^{\mu\nu}
+ \frac16\,\square{\hat P} \Big\}\,,
\label{gr-formula}
\eeq
where $\,\om\,$ is the parameter of
dimensional regularization and
$$
{\hat P}\,=\,{\hat \Pi}\,+\,\frac{\hat 1}{6}\,R\,\,
\qquad \mbox{and} \qquad
\hat{S}_{\mu\nu} \,=\,[\na_\mu\,,\,\na_\nu]\,,
$$
where the commutator of covariant derivatives is calculated 
in the corresponding vector space. Using the formula
(\ref{gr-formula}) we obtain
$$
\left.
\frac{i}{2}\,\Tr \ln \hat{\cal H}_h
\right|_{div}\,=
\,\frac{1}{2(4\pi)^2(2-\om)}\,\int d^4x\,\sqrt{-g}\,\tr\,
\Big\{\,\frac{\hat{1}}{180}\,\Big(
\,R_{\mu\nu\al\be}^2-R_{\mu\nu}^2+\square R\,\Big)
$$
\beq
+ \frac12\,\Big(M_h^2+\frac16\,R\Big)^2
+ \frac16\,\square\Big(M_h^2+\frac16\,R\Big)\Big\}
\label{back-5}
\eeq
for the contribution of the field $\,h$.
In the second case the operators $\,\hat{1}\,$ and
$\,\hat{\Pi}\,$ have matrix form
\beq
\hat{1}\,=\,
\left\| \begin{array}{cc}\de^\nu_\mu & 0 \\
0                   & 1 \end{array}
\right\|
\qquad\mbox{and}\qquad
\hat{\Pi}\,=\,
\left\| \begin{array}{cc} - R^\nu_\mu
+ M_A^2 \de^\nu_\mu & -i\sqrt{2}\,e\,(\pa_\mu v) \\
-i\sqrt{2}\,e\,(\na^\nu v) & M_\eta^2
\end{array}\right\|\,.
\label{back 0}
\eeq
Performing calculations according to (\ref{gr-formula}),
we arrive at
\beq
\left.
\frac{i}{2}\Tr \ln \hat{\cal H}
\right|_{div}& = &
\frac{1}{2(4\pi)^2(2-\om)}\int d^4x\sqrt{-g}
\,\tr\,\Big\{\frac{1}{36}\big(
R_{\mu\nu\al\be}^2-R_{\mu\nu}^2\big)
\nonumber
\\
& + & \frac12\,R_{\mu\nu}^2-2e^2(\na v)^2
+ 2\Big(M_A^2+\frac16\,R\Big)^2
- \Big(M_A^2+\frac16\,R\Big)R
\nonumber
\\
& + & \frac12\,\Big(M_\eta^2+\frac16R\Big)^2
+ \,\frac23\,\square M_A^2
+\frac16 \,\square M_\eta^2
\Big\}.
\label{back-6}
\eeq
Finally, the bilinear form of the ghost action (\ref{3.10})
can be written as
\beq
\hat{\cal H}_{gh}
\,=\,\Box\,+\,M_{gh}^2\,,\quad\mbox{where}\quad
M_{gh}^2\,=\,m^2 \,+\, \frac{e^2}{\la}\,\xi {\cal K}\,.
\label{ghost operator}
\eeq
The ghost contribution to the divergencies have the form
$$
\left.
-\,i\,\Tr \ln \hat{\cal H}_{gh}\right|_{div}\,=
\,\frac{1}{2(4\pi)^2(2-\om)}\,\int d^4x\,\sqrt{-g}\,\tr\,
\Big\{\,-\,\frac{1}{90}\,\Big(
\,R_{\mu\nu\al\be}^2-R_{\mu\nu}^2+\square R\,\Big)
$$
\beq
- \,\Big(M_{gh}^2+\frac16\,R\Big)^2
- \frac13\,\square\Big(M_{gh}^2+\frac16\,R\Big)\Big\}\,.
\label{back-66}
\eeq
The total expression for the divergencies of the
vacuum effective action in the theory with SSB is
\beq
{\bar \Ga}^{(div)}_1
&=&\frac{1}{2(4\pi)^2(2-\om)}\int d^4x\sqrt{-g}\,
\Big\{\,\frac{1}{2}\,(3m^4+m_h^4)
\,-\,\Big(\xi-\frac16\Big)(m^2+m^2_h)\,R
\nonumber
\\
\nonumber
\\
&-&
\frac{2\,m^2}{3}\,R
+\Big(\frac{3e^2m^2}{\la}+
m^2 + 3m_h^2 \Big)\xi{\cal K}
\,+\,\frac{7}{60}C_{\mu\nu\al\be}^2
- \frac{8}{45}\,E
\nonumber
\\
\nonumber
\\
&+&
\Big(\xi-\frac16\Big)^2\,R^2
-\frac{7}{90}\Box R
+\Big(\frac{e^2}{2\la}+\frac23\Big)\xi \square{\cal K}
+\Big(\frac{3e^4}{2\la^2}+\frac{e^2}{\la}+5\Big)
\Big(\xi {\cal K} \Big)^2
\nonumber
\\
\nonumber
\\
&-&
\Big[\Big(\xi-\frac16\Big)\Big(\frac{e^2}{\la}+4\Big)
+\frac{2e^2}{3\la}\Big]\,R\cdot\xi {\cal K} - 2e^2(\na v)^2
\Big\},
\label{divergent}
\eeq
where the last term can be integrated by parts and
replaced by (neglecting the surface term)
\beq
-\,2\,e^2(\na v)^2\,\,\longrightarrow\,\,
2e^2v^2\,(\xi R -\xi {\cal K} )
\,=\,m^2 (\xi R -\xi {\cal K} )
\,+\,\frac{e^2}{\la}\,\xi {\cal K} \,(\xi R -\xi {\cal K} )\,.
\label{kinetic}
\eeq
Finally, disregarding the surface terms, we have
\beq
{\bar \Ga}^{(div)}_1
&=&\frac{1}{2(4\pi)^2(2-\om)}\int d^4x\sqrt{-g}\,
\Big\{\,\frac{1}{2}\,(3m^4+m_h^4)
\,-\,\Big(\xi-\frac16\Big)m^2_h\,R
\nonumber
\\
\nonumber
\\
&-&
\frac12\,m^2R
\,+\,\Big(\frac{e^2}{\la}\,m^2
\,+\,m_h^2 \Big)\cdot3\,\xi{\cal K}
\,+\,\frac{7}{60\,}C_{\mu\nu\al\be}^2
- \frac{8}{45}\,E
\nonumber
\\
\nonumber
\\
&+&
\Big(\xi-\frac16\Big)^2\,R^2
+ \Big(\,\frac{3e^4}{2\la^2}+5\Big)
\,\big(\xi {\cal K} \big)^2
\,-\,
\Big[\,4\,\Big(\xi-\frac16\Big)
+\frac{e^2}{2\la}\Big]\,R\cdot\xi {\cal K} \,\Big\}\,,
\label{divergent11}
\eeq

The expression above differs from what is usually expected
from the divergencies of the quantum field theory in an
external gravitational field.
Along with the usual local terms, there are many terms which
look local only when they are expressed in terms of $\,v\,$
or $\,{\cal K} $. After replacing the expressions
(\ref{SSB 4}) and (\ref{back 24}) into (\ref{divergent11}),
it becomes clear that
these terms are indeed non-local with respect
to the background metric $\,g_{\mu\nu}$.

Since the appearance of a non-local divergences is quite
a surprising result, let us explain
it in more details. The $\xi {\cal K}$-dependent terms
may be rewritten, using (\ref{back 24}), in terms of
VEV $\,v\,$ as follows (here $m^2_1$ is some dimensional
parameter)
\beq
m_1^2\xi {\cal K}=2\la m_1^2v^2-\mu_0^2 m_1^2\,,
\nonumber
\\
R\,\xi {\cal K}=2\la R\,v^2- R\mu_0^2 \,,
\nonumber
\\
\big(\xi {\cal K} \big)^2=4\la^2v^4-4\la v^2\mu^2+\mu_0^4\,.
\label{more 1}
\eeq
Thus, the following new structures emerge in the
counterterms:
\beq
v^4\,,\quad m_1^2v^2\,,\quad Rv^2\,.
\label{more 2}
\eeq
This can be compared to the expression for divergences
for an ordinary real massive scalar field $\chi$ with
quartic interaction and nonminimal coupling
\beq
\Ga^{div}_{scal}=-\frac{\mu^{n-4}}{\vp}\int d^nx\sqrt{-g}
\,\Big\{\frac{\la^2}{8}\chi^4
-\frac{\la}{2}\chi^2 m_1^2-\frac{\la}{2}\chi^2
\big(\xi-\frac16\big)R \Big\}+ (\mbox{vac. terms})\,,
\label{more 3}
\eeq
where $\chi$ is a background scalar field, independent
on the metric. Of course, this is quite similar to
(\ref{more 2}). The only difference between the
two cases is that the
VEV $\,v\,$ in (\ref{more 2}) is not independent on
metric but instead is given by the nonlocal expression
(\ref{SSB 4}). We can see that the non-localities
which appear at the classical level in (\ref{SSB 4})
emerge also in the one-loop (and of course in the
higher loops) divergences.

The positive
feature is that all necessary counterterms (local and
non-local) have the form
of the induced gravitational action (\ref{SSB 80})
plus the standard local gravitational terms such as the
cosmological constant, Einstein-Hilbert and higher
derivative terms. The important consequence is that
the theory with SSB is renormalizable in curved
space-time. However, in order to achieve renormalizability,
the corresponding non-local terms must be included
into the classical action of vacuum along with the
usual local ones \cite{birdav,book}
\beq
S_{vac,1}\,=\,-\,\int d^4x\sqrt{-g}\,
\Big\{\,a_1\,C_{\mu\nu\al\be}^2\,+\,a_2\,E
\,+\,a_3\,\square R\,+\,a_4\,R^2\,+\,a_5\,R\,+\,a_6\,\Big\}\,,
\label{local vacuum}
\eeq
Using the expression for the divergences (\ref{divergent11}),
one can establish the necessary set of the non-local terms
\beq
S_{vac,2}\,=\,-\,\int d^4x\sqrt{-g}\,
\Big\{\,q_1\,\xi {\cal K} \,+\,q_2\,R\,\xi {\cal K}
\,+\,q_3\,\Big(\xi {\cal K} \Big)^2\,\Big\}\,.
\label{non-local vacuum}
\eeq
After these terms are included, we have the total
classical action $\,S_{vac,1}+S_{vac,2}\,$, which
must be compared to the induced action of vacuum in the
theory with SSB (\ref{SSB 9}).
In the theory with the complete vacuum action
$\,S_{vac,1}+S_{vac,2}\,$, the
non-local counterterms can be removed by renormalizing
the parameters $\,q_1,\,q_2,\,q_3$. The observable
values of these parameters are defined as sums of
the induced ones plus the values of the renormalized
vacuum parameters -- the last depend on the choice of
the renormalization condition.

One may worry about the potential problems with the
non-local terms in the classical action of vacuum (and
also in the induced action of vacuum), such as unitarity
of the $S$-matrix. Let us remember that what
we obtained is a direct consequence of the SSB phenomena
in curved space. The appearance of the non-localities
looks inevitable in this framework. On the other hand,
the non-localities which we are discussing here
possess very special properties:
\vskip 2mm

$\bullet$ \quad They emerge in the action of an external
gravitational field, therefore they have nothing to do
with the unitarity of the $S$-matrix of the quantum
theory. The very concept of an external
field implies that it satisfies proper equations
of motion, with certain boundary and initial conditions.
Therefore, there is no real locality in the vacuum
sector anyway.
\vskip 2mm

$\bullet$$\bullet$ \quad
The physical effect of the non-localities is
extremely weak, at least in the finite order in the
curvature expansion. The point is that the non-localities
are related to the Green functions
$\,(\square+4\la v_0^2)^{-1}\,$ and therefore the effect
of this non-localities may be observed only at the
distances comparable to $\,(\sqrt{\la} v_0)^{-1}$.
Since, according to the experimental data
$\,\sqrt{\la} v_0 > 50\,GeV\,$ in the case of the SM
of particle physics, the non-local terms are
irrelevant in the modern Universe.

\vskip 2mm

$\bullet$$\bullet$$\bullet$ \quad The conclusion
of the previous point may be not completely safe.
One can foresee the following two exceptions:
\quad {\it i)} The resummation of the
curvature expansion series may produce the
massless-type non-localities. As far as we do not
have control over these series \footnote{These series are 
obtained by inserting expansion (\ref{SSB 4})
into (\ref{SSB 80}) plus the quantum corrections.}, 
we can not exclude
this possibility, which may lead to very interesting
cosmological consequences like the IR running of
the CC due to the remnant quantum effects of the
decoupled heavy particles \cite{rgCC}.
\quad {\it ii)} If we consider a very
light scalar with the SSB (e.g. some
version of a quintessence), then the non-localities
may become potentially observable and in particular
may lead to the slight modification of the Newton
law. Such modifications may either put restriction
on the corresponding model or become relevant
for the Dark Matter problem. This possibility is
an interesting problem which deserves special
investigation.

Using the expression for the divergences
(\ref{divergent11}), we can derive the
$\,\be$-functions for the parameters of the usual
vacuum action (\ref{local vacuum}) and for the
parameters
of the non-local action (\ref{non-local vacuum})
which are necessary in the theory with the SSB.
Let us write down the
$\,\overline{MS}$-scheme $\,\be$-functions
for those parameters of the vacuum action which do
not correspond to the surface terms. In the
expressions below we used the mass $\,m_h\,$
(because this mass appears in the non-localities
at the classical level (\ref{SSB 9})) and the
relation $\,m^2=(e^2/2\la)\cdot m_h^2$.
\beq
\be_1^{\overline{MS}}\,=\,\frac{7}{60(4\pi)^2}\,,
\qquad\quad
\be_2^{\overline{MS}}
\,=\,-\,\frac{8}{45(4\pi)^2}\,,
\qquad\quad
\be_4^{\overline{MS}}
\,=\,\frac{1}{(4\pi)^2}\,\Big(\xi-\frac16\Big)^2\,,
\label{beta 123}
\eeq
\beq
\be_5^{\overline{MS}}\,=\,-\,\frac{m^2_h}{(4\pi)^2}
\,\Big(\,\frac{e^2}{4\lambda}\,+\,\xi-\frac16\,\Big)\,,
\qquad\quad
\be_6^{\overline{MS}}\,=\,\frac{m^4_h}{(4\pi)^2}
\,\Big(\,\frac{3e^2}{2\lambda}\,+\,1\,\Big)\,,
\label{beta 56}
\eeq
\vskip 1mm
\beq
\be_{q1}^{\overline{MS}}\,=\,\frac{3\,m^2_h}{(4\pi)^2}\,
\Big(\,\frac{e^4}{2\la^2}\,+\,1\Big)\,,
\qquad\quad
\be_{q2}^{\overline{MS}}\,=\,-\,\frac{1}{(4\pi)^2}
\,\Big[\,4\Big(\xi-\frac16\Big)\,+\,\frac{e^2}{2\la}\,\Big]\,,
\nonumber
\\
\nonumber
\\
\be_{q3}^{\overline{MS}}\,=\,\frac{1}{(4\pi)^2}
\,\Big(\,\frac{3e^4}{2\la^2}\,+\,5\,\Big)\,,
\label{beta q123}
\eeq
Let us remark that the signs of the $\,\be$-functions
$\,\be_1^{\overline{MS}}$, $\,\be_2^{\overline{MS}}\,$
for the higher derivative terms in this paper are different
from the ones of \cite{apco} due to the opposite sign
of the classical action. In the next section we shall derive
the physical $\,\be$-functions in the mass-dependent
(momentum subtraction) renormalization scheme and will use
the above expressions (\ref{beta 56}) and
(\ref{beta q123}) in order to check these $\,\be$-functions
in the UV limit.

\section{Renormalization
and decoupling in the SSB theory}

In order to observe the decoupling of the massive degrees
of freedom at low energies in the theory with SSB, one has
to apply the mass-dependent scheme of renormalization in
curved space-time. The most economic way of doing this is
to perform the covariant calculation using the heat-kernel
method. The existing results for the heat kernel enable
one to perform practical calculations at the second order
in curvature \cite{Avramidi,vilk}. This method of calculation
is completely equivalent to the derivation of the
polarization operator of graviton (or vertices, in the
higher orders in curvatures) due to
the quantum effects of the matter fields \cite{apco}.

The calculation of the effective action can be mainly
performed using the results
for the massive vector and massive scalar\footnote{The
only piece which requires a special calculation
is the non-diagonal sector of the operator $\,{\hat \Pi}\,$
in (\ref{back 0}).}. The one-loop contribution to the
Euclidean effective action is given by the sum of three
terms
\beq
{\bar \Ga}^{(1)}\,=
\,-\,\frac12\,{\Tr}{\ln}\,\hat{\cal H}_h
\,-\,\frac12\,{\Tr}{\ln}\,\hat{\cal H}
\,+\,     {\Tr}{\ln}\,\hat{\cal H}_{gh}\,.
\label{one loop}
\eeq
The operators $\hat{\cal H}_{gh}$ and $\hat{\cal H}$
correspond to the fields with the mass $m$, while
the operator $\hat{\cal H}_h$ correspond to the
field with the mass $m_h$. Therefore we can use, for
each of these the three operators, the standard
Schwinger-DeWitt representation, e.g.
\beq
-\,\frac12\,{\Tr}{\ln}\,\hat{\cal H}
\,=\,-\,\frac12\,\int_{0}^{\infty}\,
\frac{ds}{s}\,e^{-sm^2}\,\tr\,K(s)\,.
\label{effective}
\eeq
An explicit expression for the heat kernel $\,K(s)\,$
of the operator $\,\hat{1}\Box+\hat{\Pi}\,$ has been
found in \cite{Avramidi,vilk}, and we can use this 
result directly for the three operators of interest.

Since the practical
calculations has been described in the second reference
of \cite{apco}, we will not go into details here. Let us
present the final
expression for the $\,{\cal O}(R^2)$-terms in the
effective action
$$
{\bar \Ga}^{(1)}
\,=\,\frac{1}{2(4\pi)^2}\,\int d^4x \,\sqrt{-g}\,
\left\{\,\frac{3m^4+m_h^4}{2}\cdot\Big(\frac{1}{2-w}
+\frac32\Big)
\right.
$$$$
\left.
\,+\,
\frac{3m^4}{2}\ln \big(\frac{4\pi \mu^2}{m^2}\big)
\,+\,
\frac{m_h^4}{2}\ln \big(\frac{4\pi \mu^2}{m_h^2}\big)
\right.
$$$$
\left.
+\,\Big[\,\Big(\frac{3e^2}{\la}+1\Big)\,\xi{\cal K}
\,-\,\Big(\xi +\frac12\Big)\,R
\,\Big]\cdot m^2
\Big[\,\frac{1}{2-w}
+\ln \big(\frac{4\pi \mu^2}{m^2}\big)+1\,\Big]
\right.
$$$$
\left.
+\,\Big[\,3 \xi{\cal K}\,-\,\Big(\,\xi-\frac16 \Big) R\,\Big]
\cdot m_h^2\Big[\,\frac{1}{2-w}
+\ln \big(\frac{4\pi \mu^2}{m_h^2}\big)+1\,\Big]
\right.
$$$$
\left.
+\,\frac12 C_{\mu\nu\al\be} \Big[\frac{7}{30(2-w)}
+\frac{13}{60}\ln \big(\frac{4\pi \mu^2}{m^2}\big)
+\frac{1}{60}\ln \big(\frac{4\pi \mu^2}{m_h^2}\big)
+ k^{total}_W(a,a_h)\Big] C^{\mu\nu\al\be}
\right.
$$
$$
\left.
+\,R \,\Big[\,\Big(\xi-\frac16\Big)^2\cdot
\Big(\,\frac{1}{2-w}
+ \ln \big(\frac{16\pi^2 \mu^4}{m^2m_h^2}\big)\,\Big)
+ k_R(a)+ k_R(a_h)+ k_R^{gv}(a)\,\Big]\,R
\right.
$$
$$
\left.
+\,R \,\Big[\,
\Big(1+\frac{3e^2}{\la}\Big)\cdot
\frac{3Aa^2-a^2-12A}{18\,a^2}\,\,+\,\,
\frac{3A_ha_h^2-a_h^2-12A_h}{6\,a_h^2}
\right.
$$
$$
\left.
-\,3\,\Big(\xi-\frac16\Big)\cdot
\Big(\frac{1}{2-w}+\ln \big(\frac{4\pi \mu^2}{m_h^2}\big)
+ 2A_h\Big)
\right.
$$
$$
\left.
-\,\Big(\,\big(1+\frac{e^2}{\la}\big)\,\big(\xi-\frac16\big)
\,+\,\frac{2e^2}{3\,\la}\,\Big)\cdot
\Big(\,\frac{1}{2-w}+\ln \big(\frac{4\pi \mu^2}{m_h^2}\big)
+ 2A\Big)\,\Big]\,\xi{\cal K}
\right.
$$
$$
\left.
+\,\xi{\cal K} \,\Big[\,
\Big(\frac{3e^4}{2\la^2} + \frac{e^2}{\la} + \frac12\Big)\cdot
\Big(\,\frac{1}{2-w}+\ln \big(\frac{4\pi \mu^2}{m^2}\big)
+ 2A\Big)
\right.
$$
$$
\left.
+\,\frac92\,
\Big(\,\frac{1}{2-w}+\ln \big(\frac{4\pi \mu^2}{m_h^2}\big)
+ 2A_h\,\Big)\,\Big]\,\xi{\cal K}
\right.
$$
\beq
\left.
-\,2e^2\,(\na_\mu v)\,
\Big[\,\frac{1}{2-w}+\ln \big(\frac{4\pi \mu^2}{m^2}\big)
+ 2A\Big] \,(\na^\mu v)\,\right\}\,,
\label{final}
\eeq
where
\beq
A = A(a) = 1-\frac{1}{a}\ln \frac{1+a/2}{1-a/2}
\,\,\,\,\,\,\,\,\,\,\,\,\, {\rm and} \,\,\,\,\,\,\,\,\,\,\,\,\,
a^2=a^2(m)=\frac{4\Box}{\Box-4m^2}\,.
\label{A}
\eeq
Of course, $\,A_h=A(a_h)\,$ and $\,a_h^2=a^2(m_h)$.
In the expression (\ref{final}) we used the following
notations for the formfactors:
\beq
k^{total}_W(a,a_h)\, = \,
\frac{8A_h}{15 a_h^4}\,+\,\frac{2}{45 a_h^2}
\,+\,A\,+\,
\frac{8A}{5 a^4}\,-\,\frac{8A}{3a^2}
\,+\,\frac{2}{15 a^2}\,-\,\frac{88}{450}\,,
\label{k-W}
\eeq
\beq
k_R(a)\, =
\,A\Big(\xi-\frac16\Big)^2-\frac{A}{6}\,\Big(\xi-\frac16\Big)
+\frac{2A}{3a^2}\,\Big(\xi-\frac16\Big)
+\frac{A}{9a^4}-\frac{A}{18a^2}+\frac{A}{144}+
\nonumber
\\
+\frac{1}{108\,a^2}
-\frac{7}{2160} + \frac{1}{18}\,\Big(\xi-\frac16\Big)\,,
\label{k-R}
\eeq
\beq
k^{gv}_R(a)\, =
\frac{13}{1080} - \frac{A}{24} + \frac{1}{54a^2}
+\frac{2A}{9a^4}+\frac{A}{9a^2}
\label{k-G-V}
\eeq
(this corresponds to a massive vector with an
extra compensating scalar \cite{apco}).

The divergent part of the one-loop part of the effective
action (\ref{final}) is exactly the Eq. (\ref{divergent}).
Concerning the finite part, it is easy to see that the
non-localities of the expression (\ref{final}) have two
sources. First of all we meet the tree-level non-localities
inside each $\,\xi{\cal K}$, and moreover there are
$\,a$-dependent and $\,A$-dependent non-localities which
have the structure similar to the one for the
usual massive fields \cite{apco}. If considering
the $\,\be$-functions, the first type of the non-localities
does not matter, hence one can expect to meet the same
result as in the $\,\overline{MS}$-scheme in the
high energy regime and the standard decoupling \cite{apco}
at low energies.

It is not difficult to confirm the last statement by
direct calculation.
Using the formfactors of the expression (\ref{final})
we can derive the physical $\,\be$-functions. For this
end one has to perform the subtraction at the
Euclidean momentum square $\,-\square \to p^2=M^2\,$
and then apply the receipt
\beq
\be_C = \lim_{n\to 4}\,M\,\frac{dC}{dM}
\label{beta-mass-M}
\eeq
for the effective charge $\,C$. The coincidence with the
$\,\overline{MS}$-scheme $\,\be$-function in the UV
provides an efficient verification of the calculations.

For the usual parameters of the vacuum action, 
corresponding
to the local terms, we obtain the following results:
\vskip 3mm

\noindent
1)\quad The physical
$\,\be$-functions for the classical cosmological
constant $\,a_6$, inverse Newton constant $\,a_5\,$ and
for the coefficient $\,q_1\,$ of the non-local term
$\,\xi{\cal K} \,$ are not visible in this framework,
for the reasons which were already explained above and
in \cite{apco}. Unfortunately, at this point there is
no qualitative difference between the theory were the
masses are introduced from the very beginning and
the theory with SSB.
\vskip 3mm

\noindent
2)\quad For the coefficient of the $\,C^2_{\mu\nu\al\be}$ term we
obtain, after some algebra
\beq
\be_1 \,=\, -\,\frac{1}{(4\pi)^2} \,\Big[\,
\frac{17}{90}-\frac{1}{6a^2}-\frac{a^2}{16}
+\frac{(a^2-4)(a^4-8a^2+8)\,A}{16a^4} +
 \frac{3A_h(a_h^2 - 4) - a_h^2}{18\,a_h^4}\,\Big]\,,
\label{beta-mass}
\eeq
that is the general result for the one-loop $\be$-function,
valid at any scale. In the high energy UV limit
$\,p^2 \gg m_h^2\,$ we obtain
\beq
\be_1^{UV}\,=\,\frac{7}{60\,(4\pi)^2}\,
\,+\, {\cal O}\Big(\frac{m_h^2}{p^2}\Big)\,.
\label{W-beta-UV}
\eeq
that agrees with the $\overline{\rm MS}$-scheme result
(\ref{beta 123}).
In the IR limit $\,p^2 \ll m^2\,$ we meet
\beq
\be_1^{IR}\,=\,\frac{3}{112\,(4\pi)^2}\,
\Big(\frac{2\la}{e^2}\,+\,\frac{1}{45}\Big)\,\cdot\,
\frac{p^2}{m_h^2}
\,\,+ \,\,{\cal O}\Big(\frac{p^4}{m_h^4}\Big)\,.
\label{beta-IR}
\eeq
We have found that the IR limit of the $\,\be_1\,$ in the
theory with SSB demonstrates the decoupling, similar to
the simple massive theory \cite{apco}. There is a weak
dependence on the parameter $\,{\la}/{e^2}\,$ in the IR,
but the very fact that the decoupling occurs in the
gravitational vacuum sector of the theory with SSB does
not depend on the magnitude of the scalar coupling $\,\la$.
\vskip 3mm

\noindent
3)\quad The overall $\,\be_4$-function has the following form:
$$
\be_4=-\frac{1}{8\,(4\pi)^2}\,\Big[
\,\frac{( 4 - a^2 )}{144\,a^4}\,(5\,a^4\,A - 20\,a^2
- 5\,a^4 - 240\,A - 24\,a^2\,A)
$$
\vskip 0.5mm
$$
+ \frac{(4-a^2) \,( a^2\,A - a^2 - 12\,A)}{6\,a^2}
\,\big(\xi-\frac16\big)+
\frac{\left( 4 - {a_h}^2 \right) \,
 ({a_h}^2\,A_h - a_h^2 - 12\,A_h)}{6\,{a_h}^2}\,\big(\xi-\frac16\big)
$$
\beq
+ \big( {a_h}^2\,A_h -{a_h}^2 - 4\,A_h
\,\,+\,\, a^2\,A - a^2 - 4\,A\big)\cdot\big(\xi-\frac16\big)^2
\Big]\,.
\label{beta 4}
\eeq
Here we did not separate the terms proportional to
$\big(\xi-1/6\big)$ into $\be_3$,
as we did in the previous publication
\cite{apco}. Indeed, such distinction can be done in case
it is necessary, then we shall have an independent
$\be_3$-function. But, at the moment, our main interest is
the interface between the UV and IR limits in the Effective
Action and there is no need to enter into these details.

In the UV limit $\,a\to 2$, $\,a_h\to 2\,$ we
meet
\beq
\be_4^{UV}
\,=\,\frac{1}{(4\pi)^2}\,\Big(\xi-\frac16\Big)^2\,,
\label{beta 4 UV}
\eeq
that is exactly $\,\be_4^{\overline{MS}}\,$ from
the Eq. (\ref{beta 123}).

Indeed, we are most interested in the IR limit
$\,a\to 0$, $\,a_h\to 0$ of the $\be_4$-function
\beq
\be_4^{IR}\,=\,\frac{1}{(4\pi)^2}
\Big[\frac{11\,\la}{3780\,e^2}
+\frac{1}{180}\big(1+\frac{2\la}{e^2}\big)
\cdot\big(\xi-\frac16\big)
\nonumber
\\
-\, \frac{1}{12}\big(1+\frac{2\la}{e^2}\big)
\cdot\big(\xi-\frac16\big)^2\Big]\,\frac{p^2}{m_h^2}
\, + \, {\cal O}\Big(\frac{p^4}{m_h^4}\Big)\,,
\label{beta 4 IR}
\eeq
demonstrating the standard quadratic form of decoupling,
similar to the $\,\be_1\,$ case.
\vskip 3mm

\noindent
4)\quad
Let us now consider the nontrivial new $\,\be$-functions
for the parameters of the non-local terms
(\ref{non-local vacuum}). The $\,\be_{q2}$-function has
the form
$$
\be_{q2}=\frac{1}{4\,(4\pi)^2}
\Big[\,\Big( a^2A - a^2 - 4A \,+\, 3a_h^2A_h - 3a_h^2
- 16A_h \Big)\cdot
\big(\xi-\frac16\big)\,+\, \frac{ a^2(A-1) ( 3 e^2 - 
\lambda)}{12\,\lambda}
$$
\beq
-\, \frac{4A( 3\,e^2 + \lambda)}{a^2\lambda}
+ \frac{(2A-1)e^2}{\lambda}
+ \frac{(4A - 1)}{3}
+ \frac{(a_h^2-4)(A_ha_h^2-a_h^2-12A_h)}{4 a_h^2}
\Big].
\label{beta q2}
\eeq

The UV limit shows perfect correspondence with the
$\,\overline{MS}$-scheme expression (\ref{beta q123})
\beq
\be_{q2}^{UV}\,=\,-\,\frac{1}{(4\pi)^2}\,
\Big[\frac{e^2}{2\,\lambda} + 4\,\big(\xi-\frac16\big)
\,\Big]\,,
\label{beta q2 UV}
\eeq
while in the IR limit we meet usual decoupling
\beq
\be_{q2}^{IR}\,=\,-\,\frac{1}{(4\pi)^2}\,
\Big[\,\frac{\la}{90\,e^2}\,-\,\frac{7}{60}
\,-\,\Big(\frac12 + \frac{\la}{3e^2}\Big)
\,\big(\xi-\frac16\big)\,\Big]\,
\frac{p^2}{m_h^2}
+{\cal O}\Big(\frac{p^4}{m_h^4}\Big)\,.
\label{beta q2 IR}
\eeq
\vskip 2mm

\noindent
5)\quad
The $\,\be_{q3}$-function has the form
\beq
\be_{q3}=\frac{1}{8\,(4\pi)^2}
\Big[ \,9\left({a_h}^2 + 4\,A_h - {a_h}^2\,A_h \right)
\,\,+\,\, \Big(1+\frac{3\,e^4}{\la^2}\Big)
\,( a^2 + 4\,A - a^2\,A )
\Big]\,.
\label{beta q3}
\eeq
In the UV limit we meet correspondence with the
$\,\overline{MS}$-scheme $\,\be_{q3}$-function
(\ref{beta q123})
\beq
\be_{q3}^{UV}\,=\,\frac{1}{(4\pi)^2}
\Big(5 + \frac{3\,e^4}{2\,{\lambda}^2}\Big)\,,
\label{beta q3 UV}
\eeq
while in the IR limit there is usual decoupling
\beq
\be_{q3}^{IR}\,=\,\frac{1}{(4\pi)^2}
\Big(\frac{\la}{6\,e^2}\,+\,\frac34
\,+\,\frac{e^2}{\la}\,\Big)\,
\frac{p^2}{m_h^2}
+{\cal O}\Big(\frac{p^4}{m_h^4}\Big)\,.
\label{beta q3 IR}
\eeq

Thus, the IR behaviour of the new vacuum parameters
$\,q_2\,$ and $\,q_3\,$ in the higher derivative sector
of the theory is very similar to the one for the ``old''
parameters $\,a_1\,$ and $\,a_4$. In all cases we meet
soft quadratic decoupling when the energy-momentum
parameter of the linearized gravity $\,p^2\,$ becomes
much smaller than the masses of the particles induced
by SSB.


\section{Conclusions}

We have considered the quantum fields theory with SSB in
an external gravitational field. The SSB produces non-local
terms in the induced action of vacuum already at the 
classical level - the phenomenon which
was not, up to our knowledge, described before in the
literature. The non-local terms emerge due to the coordinate
dependence of the curvature scalar and do not show up in
the spaces of constant curvature, or in the theory with
the minimal coupling between scalar and curvature.
The appearance of the non-localities does not break such
important
properties of the quantum fields theory in an external
gravitational field, as unitarity and renormalizability.
The unitarity is preserved in the matter sector, because
the non-local terms emerge only in the action of external
gravitational field. Qualitatively, this situation is not
very much different from the theory without the non-local
terms, because in all cases the external metric satisfies
some equation of motion, depends on the boundary conditions.
Hence, the non-local effects are present even if they do
not explicitly show up in the action. Furthermore, the
physical
effects of the non-localities can be seen only at the
very short distances and are probably unobservable.
However, from the formal point of view, in the theory with
SSB one has to include the non-local terms
into the classical action of vacuum in order to provide
the renormalizability of the theory.

The physical effects of the new non-local terms do not look
very important, at least in the framework of the linearized
gravity and well-established physical theories. The reason
is that the non-localities enter the
action through the insertion of the Green functions
corresponding to the scalar particle with the mass
$\,m_h=2\sqrt{\la}v_0$, where $\,v_0\,$ is the VEV
for the flat space theory. In the case of the
Standard Model, this mass has the order of magnitude
around $\,100\,GeV\,$, and of course the effect of
the non-localities becomes significant only at very
small distances. Therefore, the effect of the non-local
terms in the recent universe can not be seen, these 
terms may be important only in the earliest periods
in the history of the universe. 
Moreover, in order to achieve these small distances
one needs to use very high energies. Then the symmetry
should be restored because the temperature of the
radiation interacting with the quantum fields is always
much greater than the energy of the gravitational
quantas. As far as the symmetry gets restored, the
induced non-local terms do not show up.

The situation may be quite different for the very
light fields, like e.g. quintessence (one of
candidates for the role of a time-dependent Dark
Energy). The quintessence is supposed to be an extremely
light field and therefore, if its mass is due to the
SSB, it should produce the non-local effects which may
be observable.

The most important result of our work is that the
decoupling really takes place for the theories with
SSB in curved space-time. We have investigated the
renormalization of both ``old'' and ``new'' vacuum
parameters in the theory with SSB and found that, in
the low-energy limit, they all vanish quadratically,
in accordance with the Appelquist and Carazzone theorem.
In this respect, the vacuum sector of the theory with
SSB is different from the matter sector, (see, e.g.
\cite{Collins}) where the
decoupling does not take place.
\vskip 8mm

\noindent
{\bf Acknowledgments.} 
The work of the authors has been partially supported by the 
research 
grant from FAPEMIG and by the fellowship from CNPq (I.Sh.). 
E.G. thanks Departamento 
de F\'{\i}sica at the Universidade Federal de Juiz de Fora
for kind hospitality. I.Sh. is grateful to the Erwin 
Schr\"odinger Institute for Mathematical Physics 
in Vienna and to the Departamento de F\'{\i}sica Te\'orica 
de Universidad de Zaragoza, where part of the work has
been done, for kind hospitality and partial financial 
support.


\end{document}